\documentclass[epj]{svjour}
\usepackage{latexsym}

\newcommand{\beq}{\begin{equation}}
\newcommand{\eeq}{\end{equation}}

\newcommand{\as}{\alpha_s} 
\newcommand{\un}{\underline}

\usepackage{graphics}
\usepackage{epstopdf}
\begin{document}
\title{Open and hidden charm production in pA collisions}
\author{Kirill Tuchin\thanks{This research was supported by the U.S. Department of Energy under Grant No. DE-AC02-98CH10886. BNL preprint number BNL-NT-05/10}
}                     
\institute{Nuclear Theory Group,\\ Physics Department, \\ Brookhaven National Laboratory,\\
Upton, NY 11973-5000, USA}
\date{Received: date / Revised version: date}
%
\abstract{
We discuss production of charmed mesons and $J/\Psi$ in p(d)A collisions at high energies. We argue that when the saturation scale $Q_s$ characterizing the parton density in a nucleus exceeds the quark mass $m$ the naive perturbation theory breaks down. Consequently, we calculate a process of heavy quark production in both open and hidden channels in the framework of the parton saturation model (Color Glass Condensate).  We demonstrate that at RHIC  such description is in agreement with experimental data on charm production.
} 
\maketitle

\section{Coherent charm production at high energies}\label{intro}

Production of a pair of a quark $q$ and antiqiuark $\bar q$  at high energies is characterized by two time scales: production time $\tau_P$ and  interaction time $\tau_\mathrm{int}$.  In pA collisions in the center-of-mass frame of the $q\bar q$ pair the production time is $\tau_P\simeq 1/(2m)$, where 
$m$ is a quark's mass. In the nucleus rest frame, this time is Lorentz time-dilated by $E_g/(2m)$ where $E_g$ is energy of a gluon in a proton from which the $q\bar q$ originates. In  terms of the Bjorken variable $x_2=(m_T/\sqrt{s}) e^{-y}$ the production time is $\tau_P\simeq1/(2 M x_2)$, where  
$M$ is a nucleon mass. At RHIC this corresponds to the production time $\tau_P\simeq 15e^y$~fm. On the other hand, the interaction time is $\tau_\mathrm{int}\simeq R_A\simeq 7$~fm.  Therefore, we conclude that in the kinematic region $y>0$ at RHIC charm is coherently produced by the nuclear color field. 

In high energy QCD coherent color field $E$ of a heavy nucleus is described by the Color Glass Condensate \cite{GLR,Mueller:wy,Blaizot:nc,MV,YuK,jmwlk,ILM,Levin:1999mw,KLN}. Being a classical field it scales as $\sim 1/g$ with the coupling. Also, the property of geometric scaling implies that the only dimensional parameter in the classical regime of QCD is the saturation scale $Q_s$. Thus, $E\sim Q_s^2/g$. The saturation scale increases as $Q_s\sim A^{1/6}e^{\lambda y/2}s^{\lambda/4}$ with atomic number and energy. When energy accumulated by the color field at the quark's Compton wavelength $g E \lambda$  becomes larger than its rest energy $m$, the color field starts coherent production of $q\bar q$ pairs out of the Dirac sea with the probability \cite{Schwinger}:
\begin{equation}\label{sch}
w\propto e^{-\frac{\pi m^2}{gE}}\,.
\end{equation} 
Thus, coherent production of heavy quarks by the Color Glass Condensate is effective when $Q_s>m$. 
Analysis of a large set of experimental data on the high energy hadronic reactions suggest that the saturation scale is $Q_s\approx 1.4 e^{0.15y}$~GeV in the center of the Gold nucleus. The charmed quark mass is $m_c\approx 1.3$~GeV. Therefore, the coherent production mechanism switches on at $y>0$ and becomes dominant at $y\simeq 2-3$ in agreement with our previous estimate.

In the next two sections I am going to discuss the phenomenological implications of the coherent charm production in open and hidden channels.

\section{Open charm production}\label{sec:1}

In the kinematic region we are interested in $\tau_P\gg \tau_\mathrm{int}$ the quark production can 
be written in a factorized form as a convolution of the valence quark 
and gluon wave functions with the rescattering factor.  In a quasi-classical approximation the differential cross section for the quark production is pA collisions reads   \cite{KopTar,Tuchin:2004rb,Blaizot:2004wv}
\begin{eqnarray}\label{main}
&&\frac{d\sigma}{d^2k \,dy}=
\int d^2b\, \int d^2x_0 \int d\alpha\,
\int \frac{d^2x\, d^2 y}{(2\pi)^3}\, e^{-i\un k\cdot (\un x-\un y)}\,\nonumber\\
&&\Phi_\mathrm{g\rightarrow q\bar q}(\un x,\un x_0,\un y, \alpha)\,
\Phi_\mathrm{q_v\rightarrow q_v\bar g}(\un x,\un x_0,\un y, \alpha)\,  \bigg(
e^{-\frac{1}{4}\, \frac{C_F}{N_c}\, (\un x-\un y)^2\, Q_s^2}\nonumber\\
&&-
e^{-\frac{1}{4}\, \frac{C_F}{N_c}\, (\un x-\un x_0) ^2\, Q_s^2}-
e^{-\frac{1}{4}\, \frac{C_F}{N_c}\, (\un y-\un x_0)^2\, Q_s^2}+1\bigg)\,,
\end{eqnarray}
where I assumed for simplicity that the dominant contribution comes from 
interaction of the $q\bar q$ pair with the target, while rescatterings of 
the gluon and the valence quark are neglected. However in general, they 
must be taken into account as well. In Eq.~(\ref{main}) I used the 
following 
notations: 
\begin{eqnarray}\label{wf1}
&&\Phi_\mathrm{q_v\rightarrow
q_v\bar g}(\un x,\un x_0,\un y, \alpha)=\nonumber\\
&&\frac{\as\, C_F}{\pi^2}\,
\frac{(\alpha\un x+(1-\alpha)\un x_0) \cdot(\alpha\un y+(1-\alpha)\un x_0)}
{(\alpha\un x+(1-\alpha)\un x_0)^2\,(\alpha\un y+(1-\alpha)\un x_0)^2}\,,
\end{eqnarray}
\begin{eqnarray}\label{wf2}
&&\Phi_{g\rightarrow q\bar q}(\un z,\un x, \un x_0, \alpha)=\nonumber\\
&&\frac{\as}{\pi}\, m^2\,\bigg(\, \frac{(\un x-\un x_0)\cdot
(\un y -\un x_0)}{|\un x-\un x_0|\, |\un y -\un x_0|}   
K_1(|\un x-\un x_0|\,m)\,K_1(|\un y -\un x_0|\,m)\nonumber\\
&&\times[\,\alpha^2\,+\,(1-\alpha)^2\,]
+\,K_0(|\un x-\un x_0|\,m)K_0(|\un y -\un x_0|\,m) \,\bigg)\,,
\end{eqnarray}
where $Q_s$ is the saturation scale, $m$ is a quark mass. 

\begin{figure}
\resizebox{0.55\textwidth}{!}{
  \includegraphics{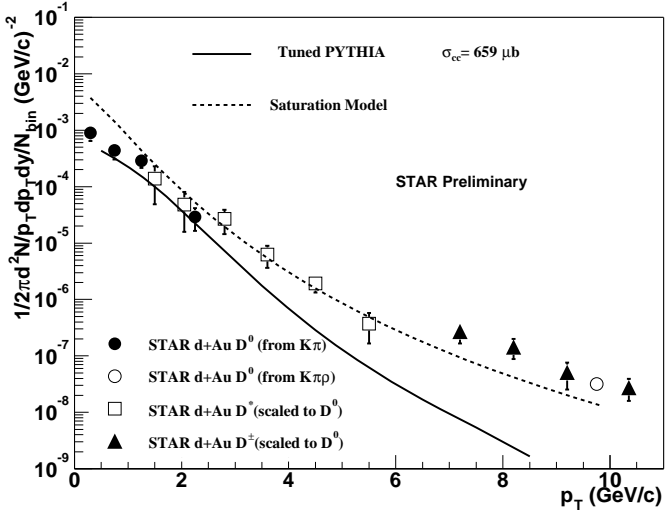}
}
\caption{Charmed meson production in dA collisions at $\sqrt{s}=200$~GeV and $y=0$ at RHIC \cite{Tai:2004bf}.  Solid line: PYTHIA event generator neglecting effect of coherence, dashed line:  prediction of a model based on gluon saturation/Color Glass Condensate \cite{Kharzeev:2003sk}. }
\label{fig:1}      
\end{figure}

The effect of the high energy evolution of the dipole-nucleus scattering amplitude can be taken into account by evolving the Glauber-Mueller  scattering amplitude $1-e^{-\frac{1}{4}\un x^2 Q_s^2}$ according to the Kovchegov equation \cite{Balitsky:1995ub,Kovchegov:1999yj}.  In practice, since the exact analytical solution to the Kovchegov equation is not known, one uses models which are supposed to satisfy the main properties of that equation. Following one such model suggested in \cite{Kharzeev:2003sk}  we predicted the spectrum of charmed mesons in dA collisions, see Figure~\ref{fig:1}.
Our prediction \cite{Kharzeev:2003sk} describes the experimental data well at transverse momenta $p_T>1$~GeV while the PYTHIA event generator based on incoherent charm production fails.  Moreover, fit of the function $A(1+p_T/p_0)^{-n}$ to the experimental data reveals that the intrinsic transverse momentum $p_0\approx 1.32$~GeV \cite{Tai:2004bf} is quite close to the saturation scale is $Q_s\approx 1.4$~GeV which indicates consistency of our approach. 

\begin{figure}
\resizebox{0.55\textwidth}{!}{
  \includegraphics{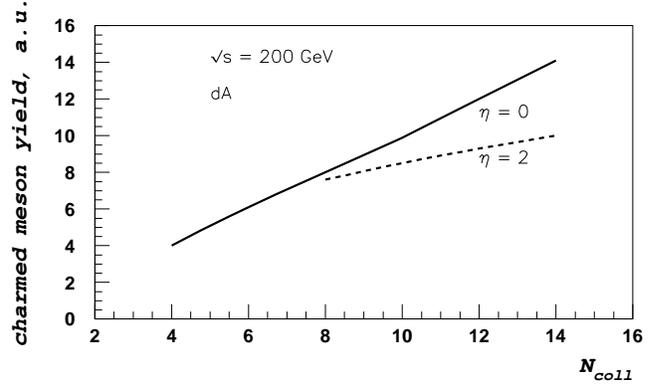}
}
\caption{ Charmed meson yield: one-particle inclusive cross section integrated over the experimental acceptance region in $p_T$ and averaged over small bins in rapidity $y$. \cite{Kharzeev:2003sk}. }
\label{fig:2}      
\end{figure}

At high energies gluon density in a nucleus of atomic number $A$ is $A^{1/3}$ times larger than one in a proton. Therefore,  it makes sense to study the centrality dependence of charmed meson yield in pA collisions. The centrality dependence is usually expressed in terms of the ``number of binary collisions"   $N_\mathrm{coll}$ experienced by the colliding systems of nucleons at given impact parameter.  The result of the calculation is presented in Figure~\ref{fig:2}\cite{Kharzeev:2003sk}.
One observes a significant coherence effect in the charmed meson production at forward rapidities. Similar result was also obtained for the charmed meson production in AA collisions \cite{Kharzeev:2003sk} although in that case the charmed meson yield is additionally suppressed by absorption in a hot nuclear medium.
 
\section{$J/\Psi$ production}
In addition to the two time scales characterizing the $q\bar q$ pair production $\tau_P$ and $\tau_\mathrm{int}$, the hidden charm production  involves another important times scale $\tau_F$ which is the time over which the charmonium bound state is formed. In the rest frame of the produced particle this time is of the order of inverse binding energy $2/(M_{\psi'}-M_\psi)$ (we concentrate on $J/\Psi$ production). In the nucleus rest frame in the RHIC kinematic regime $\tau_F\simeq 42\,e^y$ fm $\gg \tau_P$. We can therefore distinguish several kinematic regimes characterized by the relations between the time scales $\tau_P$, $\tau_\mathrm{int}$ and $\tau_F$. 

At $y>0$ the $ c\bar c$ pair is produced coherently on a whole nucleus and $J/\Psi$ pair is formed 
outside it. In that case we can calculate the $J/\Psi$ production cross section as a convolution of four time-separated amplitudes: gluon emission off a valence quark of a proton, splitting of the gluon into $c\bar c$ pair, coherent interaction of a $c\bar c$ dipole of a given transverse size with a nucleus  and subsequent formation of the $J/\Psi$ wave function.  In the quasi-classical approximation the final result is \cite{KT}:
\begin{eqnarray}\label{jpsi}
\frac{d\sigma_\psi}{dy}&=& S_A\,xG(x_1,M_\psi^2)\,\frac{3\Gamma_{ee}}{(2\pi)^2 48\alpha_{em}M_\psi}\,\nonumber\\
&&\times\int_0^\infty\,d\zeta\,\zeta^5\,K_2(\zeta)\bigg( 1-e^{-(Q_s(x_2)\zeta/2M_\psi)^4}\bigg)\,.
\end{eqnarray}
where $\Gamma_{ee}=5.26$ KeV is the leptonic width of $J/\Psi$. In derivation of (\ref{jpsi})  we used the non-relativistic approximation to the $J/\Psi$ wave function; the scattering  amplitude of the $c\bar c$ pair is calculated in large $N_c$ approximation;  valence quarks and the intermediate gluon are treated as spectators and, finally, there are parametrically small corrections due to contributions of the real part of the amplitude and the off-diagonal matrix elements.

To study the nuclear effect in inclusive observables one usually defines the nuclear modification factor
\beq\label{nmf}
R_{pA}=\frac{d\sigma^{pA}/dy}{A\,d\sigma^{pp}/dy}\,.
\eeq
In the forward kinematic region $Q_s\gg M_\psi$. In that case (\ref{jpsi}) takes form
\beq
\frac{d\sigma^{pA}}{dy}=S_A\,xG(x_1, M_\psi^2)\frac{6\Gamma_{ee}}{(2\pi)^2\alpha_{em}M_\psi}\,,
\eeq
which implies the following behavior of the nuclear modification factor
\beq
R_{pA}(J/\Psi)\sim \frac{e^{-2\lambda y}}{s^\lambda N_\mathrm{coll}}\,,\quad M_\psi\ll Q_s\,.
\eeq
It gets suppressed both as a function of energy/rapidity and centrality. At forward rapidities coherence in the nuclear wave function has similar effect on $J/\Psi$ production as in the open charm case.

Unlike in the open charm production case we expect that at rapidities $y<\sim 0$ the $J/\Psi$ inclusive cross section in pA collisions is slightly enhanced as compared to that in pp collisions (scaled by corresponding $A$).  The reason is that when $\tau_P\simeq 2-3$ fm, $J/\Psi$ is produced on a nucleus by exchange of two gluons with a different nucleons, while the same process in pp collisions goes with one gluon exchanged and another one emitted. Additional exchanged gluon brings in an additional $A^{1/3}$ enhancement of the scattering amplitude. However, at forward rapidities this effect is screened by the multiple gluon exchange which leads to suppression of nuclear modification factor as we discussed already above.  It is important to note, that although in the kinematic region of interest the production time is smaller than the interaction time, still the formation time is large  $\tau_F>\tau_\mathrm{int}$.  Due to color transparency the effect of absorption in a cold nuclear medium is only a small correction (of the order of a few percents).   The effect of enhancement occurs  in the kinematic region $M_\psi\gg Q_s$ where we can expand the exponent in (\ref{jpsi})  and derive 
\beq
R_{pA}(J/\Psi)=A^{1/3}\sim N_\mathrm{coll}^{Au}\,,\quad M_\psi>Q_s\,,
\eeq
which means that the $J/\Psi$ production is enhanced at slightly backward rapidities at RHIC and is stronger in central events than for peripheral (since otherwise  $Q_s$ exceeds $M_\psi$) . 

It should be emphasized that the prediction of enhancement  of the $J/\Psi$ production at slightly backward rapidities is a model independent statement. However, since in that kinematic region there is no clear separation of the time scales our approach (see (\ref{jpsi})) breaks down. It certainly cannot be applied at $y<-1$.

\begin{figure}
\resizebox{0.55\textwidth}{!}{
  \includegraphics{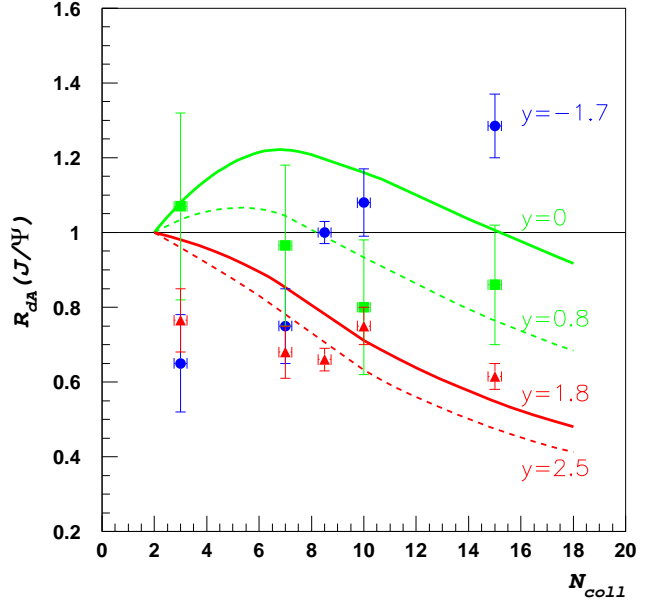}
}
\caption{  Nuclear modification factor as a function of centrality at different rapidities. Solid lines: numerical calculation \cite{KT}.  Data points (preliminary) are from \cite{deCassagnac:2004kb}.}
\label{fig:3}      
\end{figure}

As in the case effect of open charm production the high energy evolution can be taken into account 
by evolving the quasi-classical scattering amplitudes with the Kovchegov evolution equation.  In the Figure~\ref{fig:3} we present the result of our calculation \cite{KT}.

\begin{figure}
\resizebox{0.55\textwidth}{!}{
  \includegraphics{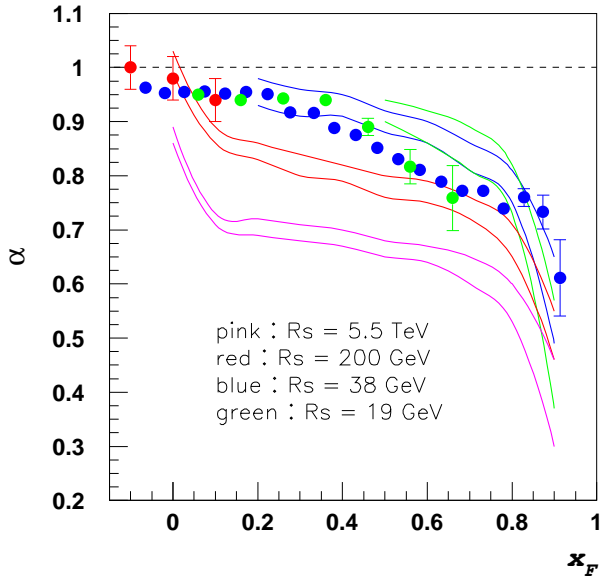}
}
\caption{  $\alpha$ as a function of $x_F$ for different energies \cite{KT}.  Data points are from \cite{Badier:1983dg,Leitch:1999ea,deCassagnac:2004kb}.}
\label{fig:4}      
\end{figure}

Another method to express the nuclear effect in the total cross sections is to define the variable $\alpha$ such that $\sigma_{pA}=A^\alpha\sigma_{pp}$. It easy to express $\alpha$  it terms of the nuclear modification factor $R_{pA}$, (\ref{nmf}). In the Figure~\ref{fig:4} we show result of our calculation together with the experimental data at different energies. Note, that since at SPS and Tevatron energies production time is smaller than the interaction time, we introduced the nuclear absorption factor $S_\psi=0.6$ \cite{Kharzeev:1996yx}.

\section{Conclusions}
We argued that if the quasi-classical coherent color field of the Color Glass Condensate is so strong that 
$Q_s\gg m$, then the production pattern of  heavy quarks with mass $m$ is the same as the light quarks. At RHIC 
at the center-of-mass energy of $\sqrt{s}=200$~GeV the saturation scale becomes much larger than the charmed quark mass at forward rapidities $y\sim 2-3$. Therefore, we expect suppression of open charm production in pA and AA as compared to pp collisions in much the same as the lighter quarks and gluons are suppressed. 

Although suppression of the nuclear modification factor for $J/\Psi$ production at forward rapidities at RHIC follows directly from the corresponding suppression of $c\bar c$ production, the detailed mechanism which triggers onset of saturation in that case is different due to a specific global structure of the $J/\Psi$ wave function. This mechanism might manifests itself in a narrow kinematic region at which $\tau_P\sim 2-3$ fm (about $y<\sim 0$) making possible $J/\Psi$ production via double gluon exchange. Experimentally, it will come about as an enhancement of the nuclear modification factor 
in peripheral and semi-central events.

Coherence effects due to saturation of the nucleus wave function will be even stronger at LHC. Since the central rapidity interval is shifted by $\ln(5.5/0.2)=3.3$ with respect to that of RHIC, we expect that the Color Glass Condensate will have a dramatic effect on particle production in general, and charm in particular, at LHC in pA and AA and collisions at $y\ge -3$, and even on pp ones at somewhat higher rapidities.


\begin{thebibliography}{50}


\bibitem{GLR}
L.~V.~Gribov, E.~M.~Levin and M.~G.~Ryskin,
Phys.\ Rept.\  {\bf 100}, 1 (1983).

\bibitem{Mueller:wy}
A.~H.~Mueller and J.~w.~Qiu,
Nucl.\ Phys.\ B {\bf 268}, 427 (1986).

\bibitem{Blaizot:nc}
J.~P.~Blaizot and A.~H.~Mueller,
Nucl.\ Phys.\ B {\bf 289}, 847 (1987).

\bibitem{MV}
L.~D.~McLerran and R.~Venugopalan,
Phys.\ Rev.\ D {\bf 49}, 2233 (1994)
Phys.\ Rev.\ D {\bf 49}, 3352 (1994)
[arXiv:hep-ph/9311205].

\bibitem{YuK} 
Y.~V.~Kovchegov,
Phys.\ Rev.\ D {\bf 54}, 5463 (1996)
Phys.\ Rev.\ D {\bf 55}, 5445 (1997)

\bibitem{jmwlk}
J.~Jalilian-Marian, A.~Kovner, A.~Leonidov and H.~Weigert,
Phys.\ Rev.\ D {\bf 59}, 014014 (1999)

\bibitem{ILM}
E.~Iancu, A.~Leonidov and L.~D.~McLerran,
Nucl.\ Phys.\ A {\bf 692}, 583 (2001)

\bibitem{Levin:1999mw}
E.~Levin and K.~Tuchin,
Nucl.\ Phys.\ B {\bf 573}, 833 (2000)

\bibitem{KLN}
D.~Kharzeev and M.~Nardi,
Phys.\ Lett.\ B {\bf 507}, 121 (2001)
D.~Kharzeev and E.~Levin,
Phys.\ Lett.\ B {\bf 523}, 79 (2001)


\bibitem{Schwinger}
J.~S.~Schwinger,
Phys.\ Rev.\  {\bf 82} (1951) 664.



\bibitem{Kharzeev:2003sk}
D.~Kharzeev and K.~Tuchin,
Nucl.\ Phys.\ A {\bf 735}, 248 (2004)




\bibitem{KopTar}
B.~Z.~Kopeliovich and A.~V.~Tarasov,
Nucl.\ Phys.\ A {\bf 710}, 180 (2002)


\bibitem{Tuchin:2004rb}
K.~Tuchin,
Phys.\ Lett.\ B {\bf 593} (2004) 66

\bibitem{Blaizot:2004wv}
J.~P.~Blaizot, F.~Gelis and R.~Venugopalan,
Nucl.\ Phys.\ A {\bf 743}, 57 (2004)

\bibitem{Balitsky:1995ub}
I.~Balitsky,
Nucl.\ Phys.\ B {\bf 463} (1996) 99

\bibitem{Kovchegov:1999yj}
Y.~V.~Kovchegov,
Phys.\ Rev.\ D {\bf 60} (1999) 034008

\bibitem{Tai:2004bf}
A.~Tai  [STAR Collaboration],
J.\ Phys.\ G {\bf 30} (2004) S809

\bibitem{KT}
D.~Kharzeev and K.~Tuchin, in preparation.

\bibitem{deCassagnac:2004kb}
R.~G.~de Cassagnac  [PHENIX Collaboration],
J.\ Phys.\ G {\bf 30} (2004) S1341

\bibitem{Kharzeev:1996yx}
D.~Kharzeev, C.~Lourenco, M.~Nardi and H.~Satz,
Z.\ Phys.\ C {\bf 74} (1997) 307

\bibitem{Badier:1983dg}
J.~Badier {\it et al.}  [NA3 Collaboration],
Z.\ Phys.\ C {\bf 20}, 101 (1983).

\bibitem{Leitch:1999ea}
M.~J.~Leitch {\it et al.}  [FNAL E866/NuSea collaboration],
Phys.\ Rev.\ Lett.\  {\bf 84} (2000) 3256








\end{thebibliography}
\end{document}